\documentclass[conference]{IEEEtran}
\pdfoutput=1 
\usepackage{listings}
\usepackage{pdfpages}
\usepackage{xcolor, multirow}
\usepackage{colortbl}
\usepackage{pgfplots}
\usepackage{subcaption,graphicx}
\usepackage{arydshln}

\begin{document}

\title{PMOs: Fast and Easy Crash Consistency for Persistent Memory}

\author{\IEEEauthorblockN{Derrick Greenspan,
Naveed Ul Mustafa, Zoran Kolega, Mark Heinrich, Yan Solihin}
\IEEEauthorblockA{
College of Engineering and Computer Science, University of Central Florida\\
Orlando, Florida, United States \\
\{derrick.greenspan, kolegazoran\}@knights.ucf.edu\\
\{heinrich, yan.solihin\}@ucf.edu \\
naveed.ul.mustafa0083@gmail.com\\
}
}
\definecolor{mGreen}{rgb}{0,0.6,0}
\definecolor{mGray}{rgb}{0.5,0.5,0.5}
\definecolor{mPurple}{rgb}{0.58,0,0.82}

\lstdefinestyle{CStyle}{
    commentstyle=\color{mGreen},
    keywordstyle=\color{magenta},
    numberstyle=\tiny\color{mGray},
    stringstyle=\color{mPurple},
    basicstyle=\footnotesize,
    breakatwhitespace=false,         
    breaklines=true,                 
    captionpos=b,                    
    keepspaces=true,                 
    numbers=left,                    
    numbersep=5pt,                  
    showspaces=false,                
    showstringspaces=false,
    showtabs=false,                  
    tabsize=2,
    language=C
}
\maketitle

\begin{abstract}
DIMM-compatible persistent memory unites memory and storage. Prior works utilize persistent memory either by combining the filesystem with direct access on memory mapped files or by managing it as a collection of objects while abolishing the POSIX abstraction. In contrast, we propose retaining the POSIX abstraction and extending it to provide support for persistent memory, using {\em Persistent Memory Objects} (PMOs). 

In this work, we design and implement PMOs, a crash-consistent abstraction for managing persistent memory.  We introduce psync, a single system call, that a programmer can use to specify crash consistency points in their code, without needing to orchestrate durability explicitly. 
When rendering data crash consistent, our design incurs a overhead of $\approx 25\%$ and  $\approx 21\%$ for parallel workloads and FileBench, respectively, compared to a system without crash consistency. Compared to NOVA-Fortis, our design provides a speedup of $\approx 1.67\times$ and $\approx 3\times$ for the two set of benchmarks, respectively.

\end{abstract}

\begin{IEEEkeywords}
Nonvolatile memory, Memory management, Operating systems
\end{IEEEkeywords}
\section{Introduction}
\label{sec:intro}
The recent arrival of DIMM-compatible Persistent Memory (PM) such as Optane DC, makes it possible for programs to host persistent data directly in memory, uniting memory and storage together while avoiding expensive serialization and deserialization.

Most proposals for persistent memory abstractions combine the filesystem with direct access (DAX) on memory mapped files. While this approach allows load/store accesses on persistent memory directly without the use of system calls~\cite{PMEM,baldassin2021persistent}, realizing the full potential of persistent memory is challenging as it requires keeping both filesystem metadata and virtual memory metadata consistent for the same underlying data. 

Other proposals for persistent memory abstractions, such as libnvmmio~\cite{choi2020libnvmmio}, intercept file IO calls and convert them to DAX  mappings.  These proposals provide byte-addressable crash consistency, but they treat persistent memory as if it were sequential I/O instead of the random-access byte addressable memory that persistent memory actually is.  

Still other proposals, such as Twizzler~\cite{bittman2020twizzler,bittman2019persistent} and MERR~\cite{xu2020MERR}, organize persistent memory as a collection of objects that can  be  mapped  to the address  space  of  a  process and hold pointer-rich data structures. MERR presents a persistent memory object (PMO) abstraction but only presents one design aspect related to embedding page tables in the PMO itself. Twizzler  presents a design that abolishes the Unix file abstraction (instead providing the POSIX abstraction as an emulation layer), and replaces a typical UNIX kernel with an exo-kernel. It uses light-weight persistent pointers to refer to persistent data structures residing in objects. While this design is promising, it requires rethinking the entire OS stack. 

One important use of a persistent memory data abstraction is to host persistent data that is recoverable after a crash. Crash consistency is challenging to ensure, as it requires reasoning about the ordering of when stores become durable inpersistent memory with respect to one another, and reasoning about which group of stores should be durable together. Partial/reordered updates may cause inconsistent state in the presence of a crash or system failure~\cite{ren2015thynvm}, resulting in partially updated nodes, dangling pointers, and more problems. To protect against such partial updates, persistent memory data structures must be updated in an atomic, crash-consistent way. Twizzler and the PMO abstraction described by MERR, does not support crash-consistent updates on object-resident user-level data structures.

It is not yet clear whether crash consistency should be entirely the application's responsibility or the system providing primitives to use~\cite{baldassin2021persistent}. The former is the approach assumed in prior work such as MERR and Twizzler. Yet, in the file system world, the latter approach is provided~\cite{bovet2005understanding}, e.g. through the {\em fsync} system call, which guarantees that all prior modifications to a file are durable at the conclusion of the call.

Listing \ref{fig:listinsertTrans} illustrates an example linked list node insertion with PMDK. The code requires annotation to dereference a pointer to read or write (D\_RO and D\_RW) and to extract pmoid (TOID). Not only is it a burden to the programmer and a source of potential bugs, it also adds overhead to pointer dereferencing as it requires pointer format translation. To achieve crash consistency, code that must be executed atomically is wrapped as a transaction by bookending it with TX\_BEGIN and TX\_END. Transactional memory was designed initially for managing concurrency. Durability was added to it in PMDK, with a library orchestrating the log creation, cache line flushing, and store fences. To avoid conflicting accesses that lead to transaction aborts, transactions need to be small. In contrast, crashes are rare, hence crash consistency generally prefers large code regions. 

\begin{lstlisting}[
    caption=Linked list node insertion using  PMDK.,
    captionpos=b,
    label=fig:listinsertTrans,
    basicstyle=\fontsize{6.5}{6.5}\selectfont\ttfamily,
    %or \small or \footnotesize etc.
    ,style=cstyle,frame=single,
    morekeywords={TX_ADD_DIRECT}
 %  linebackgroundcolor={\ifnum\value{lstnumber}=4\color{lightgray}\fi 
  % \ifnum\value{lstnumber}=5\color{lightgray}\fi
  % }, 
% struct node {int data; POBJ_ENTRY(node)*next;};
%   morekeywords={D_RW,TX_ADD_DIRECT,TOID_IS_NULL,TOID,TX_NEW,TX_BEGIN,TX_END,POBJ_ROOT,layout,pmemobj_create, POBJ_ENTRY}
]
TX_BEGIN(pool) {
    TOID(struct node) *c = D_RW(root);
    D_RW(new_node)->data = data;
    TX_ADD_DIRECT(new_node);
    while(D_RW(c)->next != NULL && D_RO(c)->data < data)
        D_RW(c) = D_RO(c)->next;
    if(c->next == NULL){
        D_RW(c)->next = &D_RW(new_node);
        TX_ADD_DIRECT(c);}
    else{tmp = D_RO(c)->next;
        D_RW(c)->next = &D_RW(new_node);
        TX_ADD_DIRECT(c);
        D_RW(c->next)->next = tmp;}
}TX_END
\end{lstlisting}

We believe that memory-mapped filesystems do not allow the full potential of the use of persistent memory for hosting persistent data. However, since most software retains the traditional POSIX abstraction and programming model, we would like to retain it, but extend it to provide a newpersistent memory data abstraction. We identify \textbf{three properties} that should be met for a persistent memory data abstraction: 
 \begin{enumerate}
     \item \textbf{Crash Consistency Primitive}: Both MERR and Twizzler do not provide applications a primitive for managing crash consistency, relegating the responsibility of managing crash consistency to the application programmer or library. We believe that to be broadly applicable, apersistent memory data system should provide a simple and intuitive primitive for managing crash consistency that is lightweight. 
     \item \textbf{Simple Extension}:  Apersistent memory data abstraction should extend, rather than radically change, existing OS abstractions.  Furthermore, the programmer should not be expected to rewrite applications in a major way to benefit from thepersistent memory data abstraction, and thepersistent memory data abstraction should work on currently available hardware.
    \item \textbf{Fast}: Accessing persistent data should be fast; the PM data abstraction should not incur much slowdown on data access. Pertinent to this is how persistent pointers are handled. Fat and indirect pointers add substantial overheads when a pointer is dereferenced. In addition, accessing the persistent data should be as lightweight as possible. 
 \end{enumerate}
 
With these properties in mind, we decided that the previously described \textbf{PMO abstraction}~\cite{xu2020MERR,mustafa-seed21,solihin2019persistent,xu2020hardware} can be adopted and extended to support crash consistency. Specifically, we propose a PMO system with psync as a crash consistency primitive that the programmer can specify. Psync ensures that all modifications to a PMO after the last invocation are not durable until the psync is called, and become durable when the invocation completes. 
It is different from a transaction in that it is object-specific rather than thread-specific (stores to non-PMO data or to a different PMO are not governed by psync, they can persist ahead of psync completion). Furthermore, unlike transactions, atomicity is not guaranteed between two consecutive psyncs, and therefore, no abort and rollback are needed.

\begin{lstlisting}[
    caption=Linked list node insertion using PMOs.,
    captionpos=b,
    label=fig:listinsertPMO,
    basicstyle=\fontsize{6.5}{6.5}\selectfont\ttfamily,
    ,style=cstyle,
    frame=single,
   morekeywords={psync}
]
attach(head, 'w');
struct node *c = head;
while(c->next != NULL && c->data < data) 
    c = c->next;
if(c->next == NULL) c->next = new_node;
else{tmp = c->next; c->next = new_node;
    c->next->next = tmp;}
psync(head);
detach(head);
\end{lstlisting}

Listing \ref{fig:listinsertPMO} illustrates how the programmer may use psync. In the figure, the programmer first attaches a PMO as they would map a file. The programmer then inserts a psync call on line 6 when data has reached a consistent point, e.g. after a node is fully inserted. In a circumstance where this type of operation is performed many times (such as inserting multiple nodes into a linked list), the programmer need not call psync after every insertion, e.g. the programmer could invoke psync after every N operations (insertions, deletions, etc.), based on the expected MTTF (mean time to failure) and recovery time. The psync primitive is more natural for a programmer familiar with traditional POSIX calls. Note also that our PMO does not involve pointer format conversion/translation, and so the code does nothing special to dereference a pointer. Furthermore, multiple threads within the same process may read or write to a PMO as it would any other shared memory address. Psync should only be invoked when a process and all its threads have finished all of its writes to the PMO. Note that, if desired, transactions could still be used with psync as long as the transaction does not contain psync call, as after the transaction is complete, the programmer may invoke psync.

Overall, this paper makes the following \textbf{contributions}:
\begin{itemize}
    \item We explore the design space for Persistent Memory Objects, a persistent memory data abstraction, and discuss the trade-offs of different pointer designs and system calls. 
    \item To provide a crash-consistent primitive, we introduce a new system call, {\em psync}. It relieves a programmer from relying on transactions or using low-level flushing and fencing to manage persistency. Our primitive is easy-to-use and less error-prone, as a programmer needs to insert only a single system call to update a persistent object.
    \item We implement PMOs in the Linux kernel and evaluate our design compared to crash-consistent NOVA-Fortis and a non-crash-consistent design. We found that our approach is still fast ($\approx 25\%$ and $\approx 21\%$ overhead for two sets of benchmarks compared to a non crash consistent solution). This is $1.67\times$ and $3\times$ faster than invoking NOVA snapshots on the two workloads at the same rate. 
\end{itemize}

\section{Background and Related Work}
\label{sec:background}

This section describes the background and related work needed for understanding persistent memory objects.
\paragraph{The Persistent Memory Fabric}
DRAM scaling has recently stalled due to the untenable increase in refresh power needed to keep smaller cells charged~\cite{irds:ieee:2017}. In contrast, persistent memory is much denser and has better scaling potential. For example, recent Intel Optane DC products have shown substantially higher density than DRAM, offering a 3TB per socket capacity~\cite{apache-pass}. In many aspects, including cost per byte, read/write latency, and read/write bandwidth, persistent memory is expected to be placed in the liminal space between DRAM and SSDs, faster than block storage, but slower than volatile memory. 
Given that, we anticipate that DRAM will continue to be used when application performance is the dominating factor, while large data will be likely to continue to reside on block based storage within a traditional file system.

\subsection{Prior Persistent Memory Abstractions}
The simplest use of persistent memory is as a drop-in replacement for DRAM to benefit from its superior density and lower cost per byte. However, this would neglect its property of non-volatility. Researchers have explored several approaches to utilizing persistent memory and its non-volatility. 

\paragraph{Filesystem approaches}
Filesystem based approaches, utilizing POSIX mmap, such as ext4-dax, PMFS~\cite{dulloor2014system}, BPFS~\cite{condit2009better}, NOVA~\cite{xu2016nova}, and NOVA-Fortis~\cite{jian2017novafortis}, host a filesystem on persistent memory and provide to a process direct access (DAX) to persistent data by mapping files to its address space via memory mappings~\cite{Coburn:2011:NMP:1950365.1950380,condit2009better,dulloor2014system,PMEM,volos:aerie:eurosys:2014,Volos2011mnemosyne,xu2016nova,jian2017novafortis}. This approach inherits the overhead of filesystems in providing direct access to persistent memory resident data that is otherwise a load/store away from a process. For example, ext4-DAX incurs up to 13x overhead compared to raw persistent memory device write bandwidth~\cite{kadekodi2019splitfs}. Furthermore, filesystem approaches need to reconcile the differing semantics of filesystems and virtual memory. 

\paragraph{Persistent-object approaches}
An alternative approach is to manage persistent memory as a collection of objects that can be mapped to address space of a process, hold data structures, and can be referenced by using persistent pointers. Previous works using this approach include Twizzler~\cite{bittman2020twizzler}, PMOs~\cite{xu2020MERR} and Mosiqs~\cite{khan2020persistent}. Twizzler provides light weight persistent pointers that enable persistent objects to be relocatable in the virtual address space of a process, but it does not provide crash-consistency primitives. Instead, it outsources crash consistency support to the programmer, either as a library, or requiring the programmer to use the low-level primitives of flushing and fencing. Though this approach works, it can be tedious as a programmer always needs to ensure that flushing and fencing are done in the proper order. Any mistake can potentially leave the data structure in an inconsistent state on a system failure. Similarly, PMOs as proposed by~\cite{xu2020MERR} lack support for crash-consistent updates of persistent data structures. 

Mosiqs~\cite{khan2020persistent} is another work which manages persistent memory as a collection of objects but relies on PMDK to provide crash consistency through transactions. However, using transactions for consistency is not a good solution for two reasons: First, to reduce number of conflicts among them, smaller sized transactions are preferred. This places a limit on number of updates that can be performed on a persistent data structure before it is persisted, but this limit is not genuine as it is imposed by the semantics of transactions and not by the abstraction. Second, transaction-based consistency can only be achieved for transactional applications, limiting the use of persistent memory to such applications~\cite{ren2015thynvm}.

\subsection{Crash Consistency}

Managing crash consistency is challenging, with several possible approaches. We identify three: \textbf{system-centric}, \textbf{application-centric}, and \textbf{Persistent Memory Object}. 

\paragraph{System-centric approach} A {\em system-centric approach}, such as that in NOVA-Fortis snapshots~\cite{jian2017novafortis}, keeps persistent data recoverable from the system point of view, i.e., filesystem metadata is kept crash consistent such that all files can be recovered to some state. However, the files may recover to some recent past version, but the application cannot specify which version easily. This property works for a filesystem, but it is not an ideal property for PMOs that contain data structures. In the case of PMOs, only the application is aware of when data structures are consistent (absent of partially-updated nodes, dangling pointers, etc.). NOVA-Fortis~\cite{jian2017novafortis} takes a consistent image of the whole filesystem and not only the files that are in active use by a process. As a result, from an application's perspective, NOVA-Fortis' snapshot approach incurs high latency and overhead. 

\paragraph{Application-centric approach}
As an alternative, an {\em application-centric} approach to managing persistency allows the programmer to manage data consistency. Most prior work with this approach focused on atomic code regions (transactions) supported by hardware or software transactional memory, and added a durability guarantee~\cite{liu2017dudetm,gray1992transaction} to transactions. Transactional memory was designed primarily for providing optimistic concurrency to allow parallel execution of critical sections that did not result in conflicting accesses between threads. Any conflicting accesses across threads in which one of the accesses is a write will abort the affected transactions; which means that transactions in a transactional memory system tend to be small (e.g. up to tens of machine instructions). In contrast, persistency is meant to deal with crashes or power failures, so much larger code regions are preferred. Therefore, using transactions for persistency management is problematic as two different requirements and semantics need to be reconciled, but in some cases, the semantics may be diametrically opposed. Furthermore, most hardware-based transactional memory implementations are best effort; as a result, the programmer still must provide alternative (non-transactional) code paths to ensure progress. 

\paragraph{The Persistent Memory Object approach}
We propose a new approach for ensuring crash consistency that is both applicable to transactional and non-transactional applications, and more lightweight from an application's perspective than taking snapshots of the entire filesystem. More specifically, we equip the PMO abstraction of persistent memory with a crash consistency feature by adding the {\em psync} system call to its programming model. The  application can call psync on a specific PMO at the point where data structures are consistent; and psync will then ensure that all updates to the PMO prior to the call are made durable. With psync, the application programmer need not wrap code regions in transactions or orchestrate data movement explicitly between the volatile and non-volatile domains (e.g. via cache flushes),  nor insert ordering instructions such as {\tt SFENCE}.

\section{Design of the PMO Abstraction}
\label{sec:design}
At a conceptual level, a PMO is a system-managed persistent memory object encapsulating persistent data structures without the backing of a file~\cite{xu2020MERR}. We provide a design for the PMO abstraction, and introduce the \textbf{psync} system call, which allows the programmer to manage crash consistency of the PMO from the application. This section discusses different approaches to solving the challenges which arise in the design of PMOs, and provides for our design.

We design the PMO abstraction with the assumption that we are utilizing contemporary hardware with persistent memory, in particular, we assume that the hardware supports Asynchronous DRAM Refresh (ADR) such that data that has reached the write pending queue (WPQ) will be written to PM even in the case of power loss (i.e., any data which has left the cache lines and entered the WPQ has reached the persistence domain~\cite{rudoff2017persistent}). Given these assumptions, data cannot be considered crash consistent until it has at least entered the WPQ\footnote{We do not assume support for Extended ADR (eADR)~\cite{eadr}, where the persistence domain is extended into the CPU caches.}. We rely on instructions that move data from the cache hierarchy into the WPQ (\texttt{CLFLUSH} or \texttt{CLFLUSHOPT}) and ones that order them with respect to others (\texttt{SFENCE}). 

There are multiple challenges that come with designing a PM data abstraction system. The following subsections present the layout of our PMO system while highlighting how our design choices contribute to achieving the goals of simple interface, fast access to PMOs and crash-consistent updates.

\subsection{Layout of PMO system}\label{layout}

We envision PMOs as a contiguious region of memory. To support crash-consistent updates, each writable PMO is backed by a shadow copy with same size as the primary copy (i.e., PMO itself) (see Section \ref{sec:crashconsistent} for more details). A shadow copy is created only when an application accesses the PMO with write permission. Only written pages are allocated in the shadow. In such case, all updates are performed on the shadow, which is then synchronized with primary copy on user-specified synchronization points in the application. No shadow copy is created for read-only PMO-access. 

\begin{figure}[htbp]
    \centering
    \includegraphics[scale=0.33]{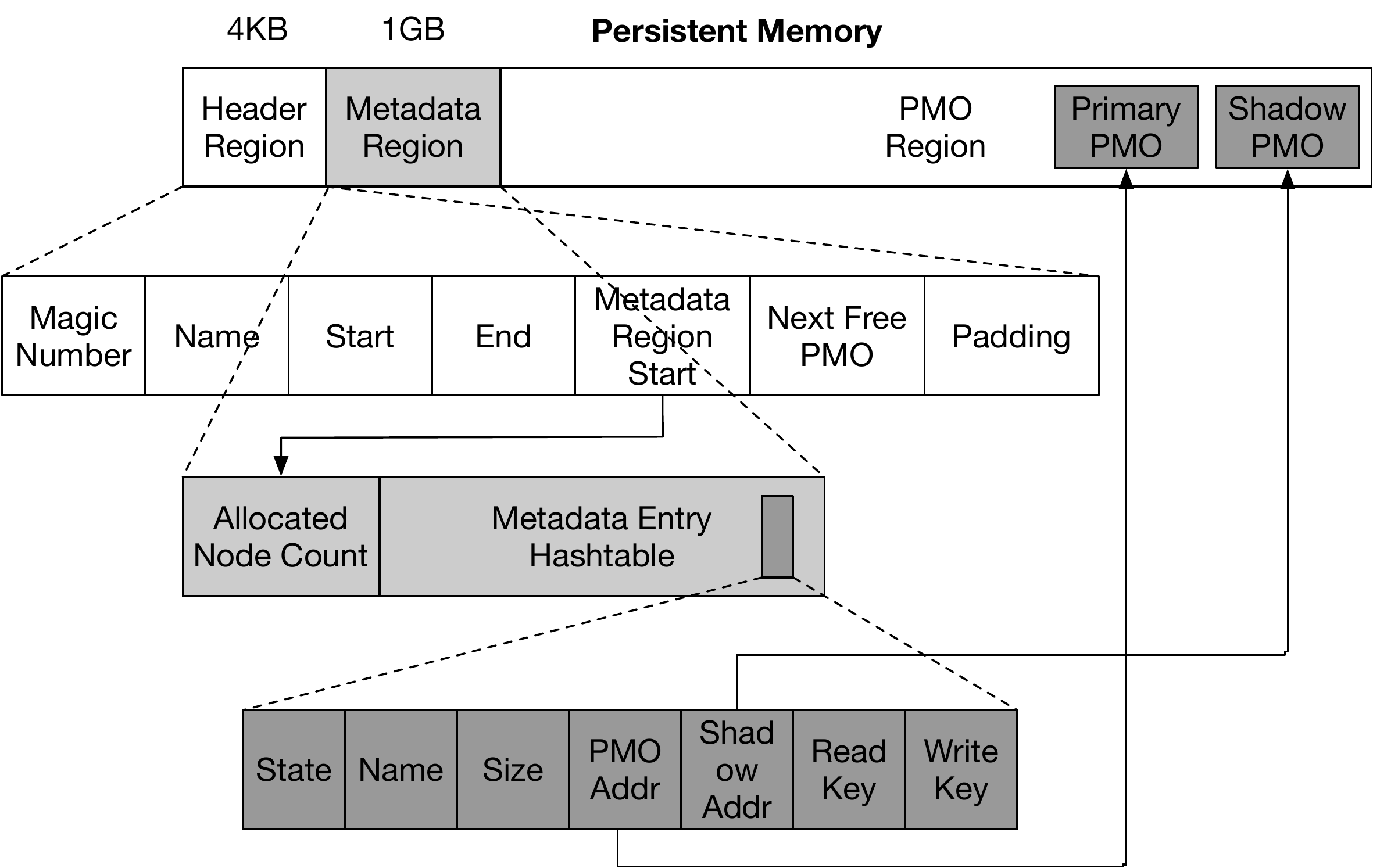}
    \caption{The layout of a PMO system.}
    \label{fig:PMOsystem}
\end{figure}

As shown in Figure~\ref{fig:PMOsystem}, our design divides the persistent memory into three regions: the {\em Header Region} which contains information important to the entire PMO system, the {\em PMO Metadata Region}, which contains a hashtable designed to make PMO operations fast, and the {\em PMO Data Region}, which contains the PMOs themselves.

Our design places the Header Region at the start of the persistent memory, making it easier for the kernel to access the header information. To keep space overhead low, we use a 4KB-size (1 page) header. It contains a magic key indicating that the device has been formatted as a PMO system, the name of the PMO system, the starting and ending addresses of the PMO system, the starting address of the PMO Metadata Region, the starting address of the next available space for a PMO, an atomic counter storing the current unique boot ID (explained below), and padding for future expansion (such as versioning information). We keep the header mapped in kernel memory for quick access.

The Metadata Region stores the metadata information of all PMOs. It consists of an 8 byte header containing the allocated node count, which represents the number of PMOs in the system. The count is used to keep track of the number of created PMOs, which is capped by the size of the hashtable.

The count is followed by the {\em Metadata Entry Hashtable}. Each hashtable metadata entry contains the minimum necessary information to ensure the correct operation of PMOs, consisting of the current state of the PMO (if it is mapped to address space of a process, if mapped with read or read/write permissions, whether or not primary copy is being synchronized with a shadow copy, and if it is being synchronized, which step in the process it is in), the name and size of the PMO, the location of the PMO (as a multiple of the page size) and its shadow copy (if any) relative to the start of the PMO Data Region, the pid and boot\_id signature (if currently attached), and the read and write access keys. To avoid false sharing, each hashtable entry should be a multiple of the size of the CPU cache line.

Finally, the vast majority of the PMO system consists of the PMO Data Region, which contains the PMOs themselves. The PMO metadata entries contain addresses that point to the PMOs and their shadows residing here. 

\subsection{Programming Interface}

\begin{table}[h]
\caption{Summary of PMO programming interface}

\centering 
{\footnotesize
\begin{tabular}{|l|l|}
\hline
\textbf{Primitive} & \textbf{Description} \\ \hline

\begin{tabular}[c]{@{}l@{}}

attach(name, perm, key)\end{tabular} & \begin{tabular}[c]{@{}l@{}}Render accessible to the calling process\\ the PMO \texttt{name}, given a matching \texttt{key} \\ with permissions \texttt{perm}.\end{tabular} 

\\ \hline

detach(addr) & \begin{tabular}[c]{@{}l@{}}Render inaccessible from the calling \\ process, the PMO pointed to by \texttt{addr}.\end{tabular} \\ \hline

psync(addr) & \begin{tabular}[c]{@{}l@{}}Force modifications to the PMO \\ associated with \texttt{addr} to be durable.\end{tabular} \\ \hline

\begin{tabular}[c]{@{}l@{}}pcreate(name,size,key)\end{tabular} & Create a PMO \texttt{name} of \texttt{size} and \texttt{key}. \\ \hline

\begin{tabular}[c]{@{}l@{}}pdestroy(name, key)\end{tabular} & \begin{tabular}[c]{@{}l@{}}Delete PMO \texttt{name}, given a matching \\ \texttt{key}, freeing the space for other PMOs.\end{tabular} \\ \hline
\end{tabular}
}
\label{tab:opvar}
\end{table}

\paragraph{pcreate}
The \textbf{pcreate}  primitive creates a PMO of a specified name, size, and key. Once created, a PMO is resident in the physical memory of the PM until destroyed. Upon invocation, the kernel searches for unoccupied space within the PMO Region by using "Next Free PMO" field of the header region (Figure \ref{fig:PMOsystem}). If requested size is larger than the remaining available space for the PMO system, then the call fails and a null pointer is returned. Otherwise, a persistent region of requested size is reserved as PMO, "Next Free PMO" field is updated and an entry is created in the hashtable before returning PMO pointer to the calling process. 

\paragraph{attach/detach}
The  \textbf{attach}  primitive maps the PMO into the virtual address (VA) space of the calling process, rendering it accessible depending on the specified permissions and key. From the perspective of the programmer, it appears that the data within the PMO has been "attached" to the VA space. Likewise, the \textbf{detach} primitive renders the PMO inaccessible from the calling process, and to the programmer, it appears that the address space associated with the PMO has been "detached" from the address space. 

If a process attempts to detach a PMO that is already detached or has never been attached, this results in undefined behavior, as the virtual address no longer has a valid mapping to the physical address of the PMO. This is very much the same effect as calling free on a pointer that has not been initialized or already freed.

\paragraph{Inter-process semantics of Attach/Detach}

\begin{figure}[htbp]
    \centering
{
\scriptsize
\begin{tabular}{llll}
\multicolumn{4}{l}{\textbf{PMOs: A(R), B(RW), C(RW)}} \\\\
\textbf{Process P1} & \textbf{Process P2} & \textbf{Process P3} & \textbf{Outcome} \\ \hdashline
attach(A, rw,) &  &  & {\color[HTML]{FE0000} invalid (permissions)} \\ \hdashline
&  attach(B,r,) &  & {\color[HTML]{036400} valid} \\ \hdashline
&  & attach(B,r,) & {\color[HTML]{036400} valid} \\ \hdashline
&  & attach(C,rw,) & {\color[HTML]{036400} valid} \\ \hdashline
& attach(C,rw,) &  & {\color[HTML]{FE0000} invalid (\textgreater 1 writer)} \\ \hdashline
attach(C,r,) &  &  & {\color[HTML]{FE0000} invalid (existing writer)} \\ \hdashline
\end{tabular}
}
    \caption{PMO inter-process sharing semantics.}
    \label{fig:attach-example} 
\end{figure}

We only allow one \textbf{process} to attach a PMO with intent to write, but allow multiple processes to attach a PMO with intent to read. Read and write permissions are {\em mutually exclusive} of each other, e.g., a PMO cannot be attached as a read by one process, and as a write by another. This avoids data consistency problems that arise from multiple writers.  Figure~\ref{fig:attach-example} illustrates this, with an example of one PMO restricted to read only access permission (PMO A), and three others allowing read and write permission (PMOs {B, C}, and {D}). If Process {P1} requests attaching  {A} with read/write access request (top line), the call returns with an error due to insufficient permissions, as {A} is restricted to read only access. Later, if Process {P2} requests attaching {B} with read only access, the attach succeeds. Process {P3}'s request attaching {B} with read only access is also granted as multiple readers are permitted. Process {P3}'s request attaching  PMO {C} with read/write access is valid, but Process {P2}'s request attaching it returns with error. Finally, an attach request for PMO {C} by Process {P1} also returns an error because there is already an existing process that has attached the writer. 

\paragraph{psync}
The \textbf{psync} primitive forces all modifications made on the shadow copy to reach a persistency domain (and thereby be rendered durable) by synchronizing it with its primary one. We design psync to have similar semantics to the POSIX \textbf{msync} and \textbf{fsync}~\cite{IEEEPOSIX}, but with only one argument: a pointer to the PMO. psync has {\em atomic semantics} for stores to the PMO, but {\em non-atomic semantics} for all other stores.

\begin{figure}[htbp]
    \centering
    \includegraphics[scale=0.32]{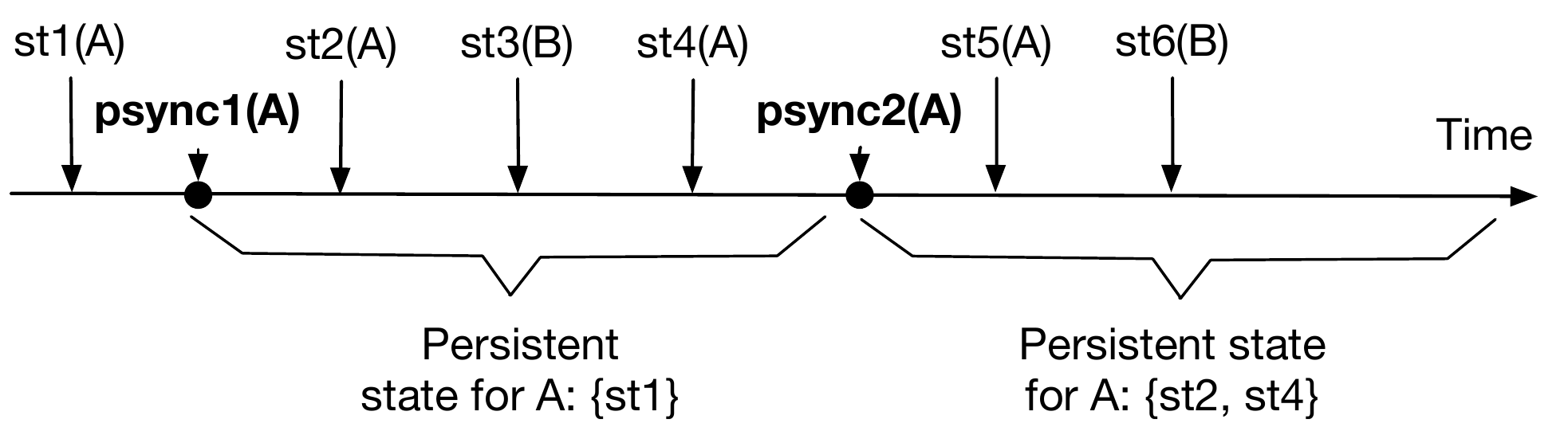}
    \caption{psync transactional semantics.}
    \label{fig:psync-example}
\end{figure}

Figure~\ref{fig:psync-example} illustrates the atomic semantics with an example, showing two psync calls for PMO {A}, with stores to A (st1, {st2}, {st4}, and {st5}) or to PMO {B} ({st3} and {st6}). If the first call completes, then the persistent state of PMO {A} in memory reflects the durable state of {st1}. Prior to the completion of the second psync, the persistent state of {PMO A} is unchanged. It is only afterwards that the persistent state of PMO {A} reflects the changes by {st2} and {st4}. How PMO B's state is affected by {st3} and {st6} is unspecified because there is no psync involving B. Therefore, the stores between the two consecutive psync calls are either entirely reflected in the persistent state or not at all. 

These data-centric semantics distinguishes psync from a transaction which is thread/code-centric. That psync's semantics prevent prior modifications from becoming persistent is also different from fsync, which does not have this guarantee. As a result, psync allows the programmer to specify crash consistent points for PMO data in their code. After a crash, the PMO state reflects the state of the most recent completed psync. From the point of view of the system, psync provides a PMO-specific persistency barrier. 

We design psync to be {\em idempotent}, where multiple invocations lead to the same outcome. This means that if psync itself is interrupted by a crash, all that needs to be done at crash recovery is to re-execute it entirely. 

\paragraph{Inter-thread semantics of psync} \label{par:interthread}
A key challenge is determining whether to prevent one thread from writing to a PMO while another thread is performing psync; and if so, how. System calls on file descriptors associated with I/O devices (such as fsync), block other I/O operations on the file until the call completes, allowing other instructions to execute during the call. However, invoking fsync on memory mapped DAX does not cause the program to block; rather, the kernel flushes each cacheline of the associated addresses; other threads may still perform operations on the mapped data. In this case, the data that are persisted to memory are crash consistent, but they are not atomic. 

As we established, psync is atomic with respect to stores to a PMO, at least in the case of single-threaded programs. However multi-threaded applications present a challenge. We identify two workable solutions: 1) either adopt fsync's model, and allow writes to a PMO being synchronized, or 2) block the entire process (and all threads) from continuing until psync has completed. Option 2) is clearly unworkable for multi-threaded programs, as there is no reason that an entire process should wait on the operations of a single PMO. Since we want psync to be an atomic operation in addition to crash consistent, Option 1) requires that unlike fsync, a case where other threads write to synchronizing PMO should result in undefined behavior, as there is no non-serializing way to prevent a thread from writing to byte-addressable memory\footnote{Unless we assume hardware specific hardware support that would allow for atomically changing the page access permissions on a range of pages, one example would be user-page protection keys (PKRU).} Consequentially, psync only ensures atomicity within threads, and requires the programmer or a library to prevent errant writes from other threads before the psync has completed. 

\begin{figure}
\centering
{
\scriptsize
\begin{tabular}{llllll}
\multicolumn{6}{l}{\textbf{PMOs: A(RW), B(RW)}}                                                                           \\
                  &                   &                                                                                     \\
\textbf{Thread 1} & \textbf{Thread 2} & \multicolumn{2}{c}{\textbf{Blocking}} & \multicolumn{2}{c}{\textbf{Non-Blocking}} \\ \hdashline
psync(A)          & psync(B)          & T1: 10          & T2: 20          & T1: 10            & T2: 10            \\ \hdashline
psync(A)          & write(B)          & T1: 20          & T2: 32          & T1: 20            & T2: 12        \\ \hdashline
psync(A)          & write(A)          & T1: 42          & T2: 52          & \multicolumn{2}{l}{\color[HTML]{FE0000} Undefined Behavior}        \\ \hdashline
\end{tabular}
}
    \caption{Blocking vs. non-blocking unit-less execution times for psync.}
    \label{fig:interprocess}
\end{figure}

Figure \ref{fig:interprocess} illustrates how multi-threaded processes affect the performance of psync, and how providing performance through non-blocking complicates the atomicity guarantee. Note that non-blocking produces correct and performant results, except that the last operation (writing and synchronizing the same PMO on separate threads) produces undefined behavior, requiring the programmer to avoid this situation. However, this requirement is worth the additional burden on the programmer, given the tremendous performance improvements possible.

\subsection{Design for Fast Access}
Latency seen by an application storing its persistent data in PMO is affected by two main factors: access latency and pointer dereferencing latency. \textit{Access latency} is the latency of creating PMOs, mapping and unmapping them in its address space, and rendering them durable in a crash-consistent manner. This latency in turn depends upon the amount of metadata that needs to managed by the operating system while controlling accesses to a PMO. \textit{Pointer dereferencing latency} is the time-overhead involved in translation between virtual PMO pointers to their physical counterparts while accessing the PMO-resident data structures. A low latency design is provided by the following design choices.

\paragraph{PMO Layout} Most filesystems use pointer chasing to locate next block of a file or track free blocks, e.g., filesystem inodes~\cite{mckusick1984fast}. Though this approach supports dynamic growth of a file, it is not conducive to fast access. Therefore, we advocate for an approach where a PMO is a contiguous region of memory with a static size set at the PMO's creation. Any PMO-resident data can be accessed by adding a given offset to base-address of the PMO; this approach is faster as it does not need to chase pointers. If the size of the data structure grows beyond what was allocated initially for the PMO, a resize operation can be performed, by creating a new PMO with a larger size and copying over the content.

\paragraph{Low-latency Attach/Detach}

In a naive approach, on invocation of an \textit{attach} system call, the kernel could map the entire PMO into the process address space, and unmap the entire PMO at \textit{detach}. This solution is expensive for a large PMO with multiple page table entries (PTEs), as PTEs need to be initialized by the kernel, invoking expensive TLB shootdowns and subsequent TLB misses. MERR~\cite{xu2020MERR} proposed embedding the page table subtree into the PMO itself, so that when a PMO is attached only one PTE needs to be initialized. As a result, this solution means that regardless of PMO size, only a single TLB shootdown is needed when a PMO is mapped into virtual memory. However, MERR's solution needs a custom hardware permission-matrix to provide access-control to PMO. Since our PMO system must work with commodity hardware, we use a different solution, {\em demand paging}~\cite{gorman2004understanding}.

When a PMO is "attached", the kernel sets a flag to indicate that future page faults for pages within a PMO should map the faulting page into a VA space. When a PMO is "detached", the kernel renders the specified address associated with a PMO inaccessible, by setting the metadata entry to the detached state, and then disabling the read/write permissions of all {\em faulted} pages, ensuring that all page faults on the address range generate a segmentation fault. This solution is not as efficient as MERR's hardware solution. However, in most cases, this solution is faster than simply mapping the entire PMO at attach time and also works on existing systems, because only accessed pages have been faulted in, instead of all of them.  
\paragraph{Low-Latency Pointer Dereferencing}

A key challenge of PMO-resident (i.e. persistent) pointers is that the VA a pointer refers to must be associated with the physical address of a PMO beyond process lifetime ~\cite{baldassin2021persistent}. This is required to ensure that pointers within and across PMOs are always valid. One solution to this challenge is to use {\em relative pointers} in object:offset format~\cite{bittman2019persistent}, and use a per-PMO Persistent Object Table (POT) for efficient pointer translation from persistent to virtual form. However, this approach adds pointer-to-VA translation on critical path to PMO access, incurring substantial latency, and increases the amount of PMO metadata. For TPC-C, persistent pointer dereferencing was reported to cause a 15\% execution time overhead~\cite{wang2017hardware}, While hardware supported pointer translation~\cite{wang2017hardware} could significantly reducing its latency, it is not clear if such a heavy-duty hardware solution is necessary. The high latency software translation is in conflict with the design goal of fast access to PMOs, while hardware based translation conflicts with the goal of PMO systems being available on current machines. 

As an alternative to relative pointers, we propose to use {\em static pointers}. Static pointers are already in VA format so just like non-persistent pointers, they can be dereferenced without additional overheads and without any need for hardware support. However, the drawbacks are that all pointers in the PMO need to be updated when the PMO mapping address changes, and that two different objects must not map to the same VA. Therefore, we adopt static pointers and assign VA range to PMOs at creation time such that no two addresses overlap. To achieve this, we use several techniques. First, to avoid an overlap between PMOs and non-persistent data, we split the effective virtual user space address space into two halves based on the most significant bit of an address: persistent and volatile, with the persistent-half reserved for PMOs and starting at (for example) the VA $x$. The kernel maps a PMO into the persistent-half of the VA space by assigning to it the address range from $x+y$ to $x+y+s$ where $y$ is the offset of the PMO in persistent memory from its start, and $s$ is its size. To prevent two PMOs from mapping to the same VA range, we assign PMOs globally unique VAs. 

Static global VA allocation for all PMO in the system incurs a risk, where if there are many large PMOs, we may run out of VA space. As a result, we advocate two mitigation strategies. First, we allocate static global VA allocation only for PMOs that are not too large. For a 48-bit address space, the persistent half can hold 128 TiB can hold 64M 2MiB-sized PMOs, or even 128K 1GiB-sized PMOs. Second, we may use static pointers for small to medium PMOs (KiBs to MiBs) but relative pointers for large PMOs (GiBs and above). 

Since the VA at which a given PMO is determined by the location of the PMO in persistent memory, our approach makes it costlier to do PMO address randomization used in MERR~\cite{xu2020MERR}; moving a PMO to a different VA requires updating all pointers in the PMO, though this is rare. However, the common case of quick dereferencing of persistent pointers {\em without} the need of software or hardware translation, or additional per-PMO metadata, makes our approach attractive.

\subsection{Design for Crash-Consistency}
\label{sec:crashconsistent}

The psync system call should persist updates in a PMO without requiring explicit logging by the programmer. Also, updates should be persisted in an atomic fashion: all or none of them should become durable. To achieve this, our PMO system manages two copies of the PMO data: the primary copy and the shadow copy. Writes to the PMO are performed to the shadow copy until psync is invoked, at which point they are copied over to the primary copy in a durable atomic manner by the system call.

As a naive approach, at the time of PMO creation, we could allocate twice of the requested PMO size in persistent memory and split it in two halves. The first half of the allocation can be used for the primary copy and the second half for the shadow copy. Since the hashtable entry for a PMO keeps track of its start address and size, calculating the starting and ending address of each copy of the allocation is simple. However, this approach is wasteful, especially when a PMO is attached only with read permissions and hence a shadow copy exists, but is never used.

Therefore, we follow a different approach where at the time of creation, a memory region of requested size is allocated and serves as the primary copy. When a PMO is attached by an application with write permission, our design allocates the size of the PMO again and designates it as the shadow copy. Our approach can potentially result in primary and shadow copies non-contiguous to each other. Therefore, we track the shadow copy through the "PMO Shadow Addr" entry pointing to location of shadow (see Figure \ref{fig:PMOsystem}). A null entry indicates that the PMO is currently detached, or attached with read only permission.

\begin{figure}[htbp]
    \centering
    \includegraphics[scale=0.55]{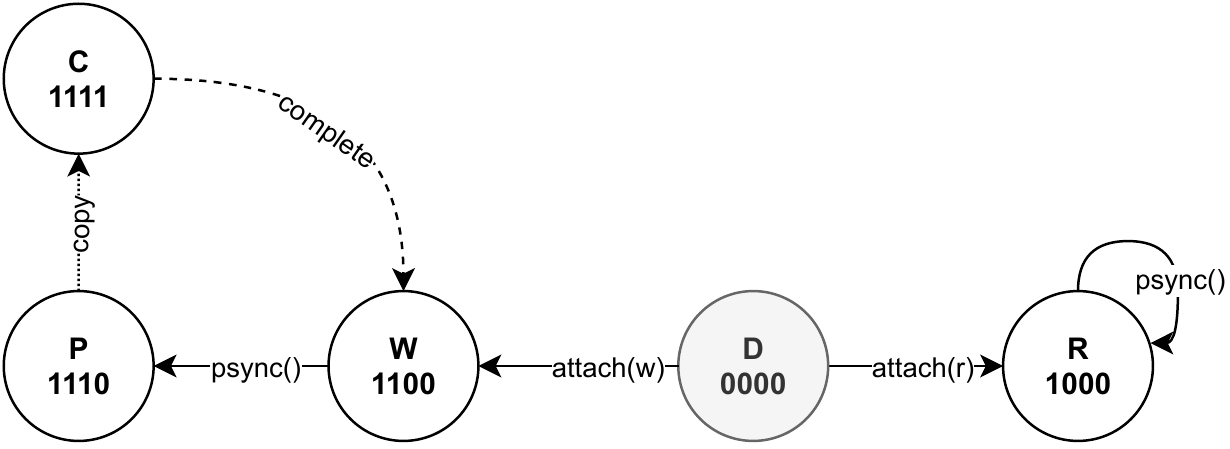}
    \caption{PMO state transitions during normal operation.} 
    \label{fig:statediagram}
\end{figure}

To achieve crash-consistent updates, under normal operation, a PMO is always in one of five states shown in Figure \ref{fig:statediagram}. The state is kept in a portion of the Metadata Hashtable that is allocated in non-cacheable pages. A state transition is performed using an atomic instruction. Since the state pages are not cacheable, the changes made by the atomic instruction also achieve durability. State D (Detached) is the initial state a PMO is created with. When attached by a process with read permission, a PMO state transitions to R (Read), where updates to the PMO are not allowed, and invocation of psync is ignored. When attached with write permissions, a PMO state transitions from D to W (Write) state where updates are permitted. If a programmer invokes psync on the attached PMO, its state transitions from W to P (Persisting), to indicate the start of psync. After the state transitions to P, the kernel performs page table walks to identify all dirty shadow pages associated with the PMO. Then, cache lines belonging to the dirty pages are flushed. After this point, the shadow copy is consistent, and the state transitions to C (Copying) to start the next step. After the state transitions to C, the kernel copies all the modifies pages from shadow to primary, flushes the cache lines, and emits a memory barrier. Only after the memory barrier completes, the state transitions from C back to W. Note that the figure shows bitvector values representing the states in our implementation, but other values are possible.

\begin{figure}[htbp]
    \centering
    \includegraphics[scale=0.55]{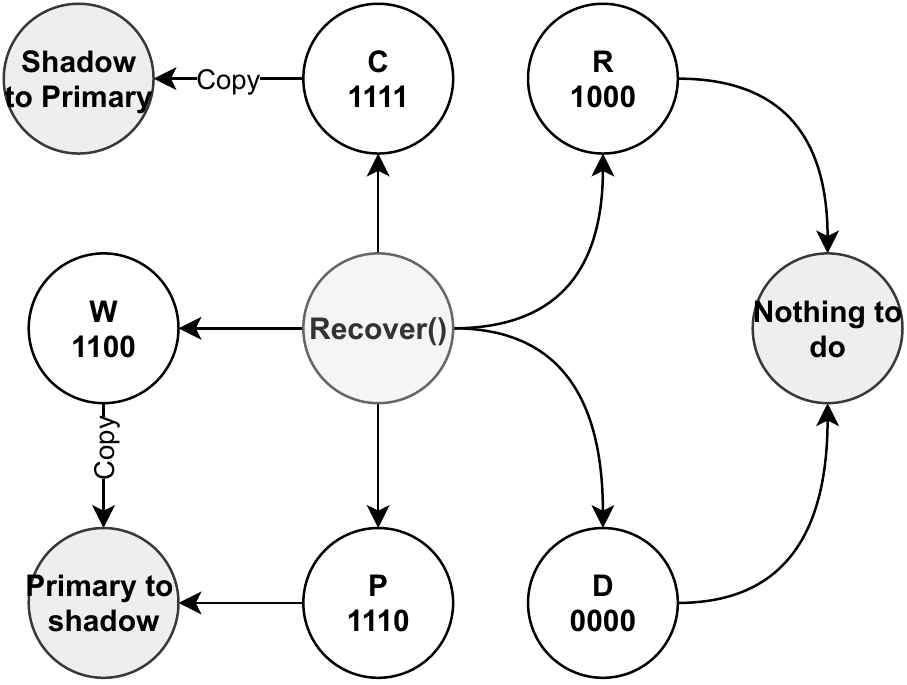}
    \caption{PMO state transitions during recovery.} 
    \label{fig:recover}
\end{figure}

\paragraph{Recovery}\label{sec:recovery}
If psync is interrupted by a crash or power failure, the kernel needs to ensure that it is recoverable. It is helpful to start with an invariant that we keep: at least one of the primary copy or shadow copy contains the consistent version of data. The recovery process depends on the state of each PMO to determine which copy to rely on as consistent, illustrated in Figure \ref{fig:statediagram}. On post-crash reboot, the kernel checks the state of each PMO. If the PMO is in the D (Detached) or R (Reading) state, then the primary and shadow PMOs are both valid, so there is nothing to do. If a PMO is in W (Writing) state, psync has not started, and the primary copy is the consistent one, so it is copied over to the shadow. This in effect removes transient updates in the shadow copy since the last psync. If a PMO is in the P (Persisting) state, psync has started but there is no guarantee that the shadow copy is consistent, so this case is treated just as the previous one for W state. Finally, if a PMO is in C (Copying) state, the shadow copy is known to be consistent and reflects all the updates until the current psync, but the primary copy might not (it could be partially copied over). In this case, the shadow copy is copied to the primary. Figure \ref{fig:recover} shows the state transition diagram for recovery.

In the case of a preceding fault, recovery takes linear O(n) time; the total time for recovery is the time it takes to copy from the shadow PMO to the primary PMO, or vice-versa, so it is bound to the size of the PMO. In the case where the PMO was in the D or R state, recovery takes O(1) time.

\section{Implementation}
\label{sec:implementation}
This section discusses our implementation of PMOs in the Linux kernel. We introduce a new data structure for maintaining persistent memory: Virtual Persistent Memory Area (VPMAs) (as a compliment to Virtual Memory Areas), the kernel modifications required to implement PMOs, the implementation of \textbf{attach}, \textbf{detach}, \textbf{psync} system calls, and our implementation of static user space virtual addresses for PMOs. 

\paragraph{Provisioning}
\label{sec:provisioning}
Intel Optane DC Memory supports namespaces which expose the region of Optane memory as a logical device~\cite{peng2019system} with different modes. We provision the namespace in devdax mode, providing direct access (DAX) to the underlying persistent memory~\cite{PMEM}. From the perspective of a program which accesses a devdax  device, the device can be directly mapped through {\em mmap}. The device is a typical {\em character device}, or stream. Analogous to block devices being formatted with \texttt{mkfs}  in Unix, we introduce a similar utility for formatting PMO systems, called \texttt{mkpmo}, that zeroes a given provisioned namespace, and then writes the PMO system data structures and information.

\subsection{Kernel Modifications}
We make several modifications to the Linux kernel, and add three new system calls, {attach}, {detach}, and {psync}. In addition, we modify the Linux per-process memory descriptor (memory management struct) and add a new red-black tree, the root of which is called \verb pmo_rb . Similar to how the \verb mm_rb  red-black tree keeps track of virtual memory areas, the \verb pmo_rb  keeps track of virtual persistent memory objects, and in particular, keeps track of the persistent memory objects associated with the process. Furthermore, we modify the user address page fault handler to support accessing PMOs through demand paging. We adopt the use of the \verb VM_SOFTDIRTY  flag as in~\cite{taskmemorychanges, emelyanovsoftdirty} to track which pages have been modified. In this way, only those pages which the process has requested access to will have their permissions modified, and only those pages which have been modified will be synchronized when psync is invoked, reducing redundant writes, and massively improving performance for larger PMOs.

\paragraph{Virtual Persistent Memory Areas}
We extend the Linux virtual memory areas (VMAs)~\cite{love2010linux,bovet2005understanding} by adding a PMO specific memory area, \textbf{V}irtual \textbf{P}ersistent \textbf{M}emory \textbf{A}reas, which we refer to as VPMAs. Our VPMA extends a VMA by containing metadata only relevant to PMOs. To implement them, we introduce a new virtual persistent memory area descriptor which we refer to as a \texttt{vpma\_area\_struct}. A VPMA contains several pointers, including a pointer to the associated VMA, a pointer to the associated PMO hashtable entry, a pointer to the entire PMO mapped in kernel memory, and a pointer to the memory descriptor of the VPMA. VPMAs also contain a read-write semaphore allowing for multiple threads to read (but only one thread to write) to the VPMA area. In addition, VPMAs contain a red-black tree containing the PIDs corresponding to the process which attached different regions of the PMO and their starting/ending address, this ensures that a thread other than the initially-attaching thread cannot unintentionally detach or synchronize the PMO.

\paragraph{User Space Addresses}
Most modern x86-64 CPUs can access only $2^{48}$ virtual addresses\footnote{With five-level page tables, Cascade Lake can access up to $2^{56}$ addresses~\cite{inteldevelopersmanual}}, and half of those virtual addresses (all addresses where the MSB is set to $1$) are kernel space virtual addresses. This means that, by default, there are $2^{47}$ user space virtual addresses and $2^{47}$ kernel space virtual addresses available. As described in the previous section, we split the user space virtual address range in half by reserving the second most significant bit (MSB) to PMO virtual user space addresses, which results in $2^{46}$ addresses for PMOs and $2^{46}$ addresses for normal user space processes. To do this in the kernel, we reduce \verb __VIRTUAL_MASK_SHIFT  from $47$ to $46$. Virtual user addresses are then assigned for all addresses below $2^{46}$, and ASLR works properly for these addresses. Since applications are given user space addresses by the kernel, this should not affect processes at all. It is possible that a program could assume that addresses above $2^{46}$ are available for it to use, but such a program would be poorly designed and non-portable anyway. Therefore, we do not believe this change breaks user space.

\paragraph{Page Fault Handler}\label{pagefaulthandler}
We modify the kernel so that when a page fault occurs in a user-space process, it will check whether the second most significant bit is asserted. If not, the kernel interprets the faulting address as a normal volatile address. Otherwise, the kernel invokes our own function, \textbf{handle\_pmo\_pagefault}. The call checks whether the VPMA is flagged as being in the detached state - if so, the permissions of the page are not touched, and the kernel generates a segmentation fault. If instead it finds that the VPMA is in the attached state, the kernel checks whether the page is mapped already; if not, the kernel maps the page to the faulted user page, sets the proper permissions and adds the address to a linked list tracking which pages have had their permissions set. In either case, the page remains mapped for the lifetime of the process; unless the PMO is reattached with different permissions, then future accesses generate a page fault and fault the new page into the process, so that the same user space address for read and write pointers can be used.

\paragraph{attach}
When an application invokes attach, the kernel attempts to determine if the PMO is currently attached to another process with {\em write permissions}, by checking the boot\_id and whether there is a PID associated with the PMO. If there is a PID associated with the PMO, and the boot\_id is the same, the kernel walks through all running process IDs to determine if any of them match the stored PID. If so, the attach returns an error. If not, the kernel attempts to determine whether or not the PMO has already been attached by the running process in the past by searching the \verb pmo_rb  for the vpma entry.  Before the kernel performs the actual process of rendering accessible to the calling process the specified PMO, it searches for metadata about the PMO in a new kernel wide radix tree (pmo\_radix\_tree). If it is not found there, the kernel searches for the PMO's hashtable entry by performing a hashtable lookup. If the kernel does not find the PMO there, then the PMO has not yet been created, and the attach call fails. Alternatively, if the kernel does find the PMO in the hashtable, then the kernel maps the PMO metadata as uncacheable via \verb ioremap_uc  \footnote{Although none of the members of the ioremap family of functions guarantee that they will emit directly accessible mappings~\cite{gorman2004understanding,bovet2005understanding}, directly accessing these mapped addresses in x86 works properly anyway.}, and then adds the PMO metadata into the radix tree for quick lookup. 

If the PMO has never been attached by the process, a VPMA is initialized. First, a pointer to the PMO hashtable entry within the VPMA (\verb pmo_ptr ) is set. Second, the kernel initializes an associated vma. Third, \verb calling_pid , a variable tracking the PID of the attaching process, is set. Finally, the newly initialized VPMA is now added into the \verb pmo_rb  red-black tree. This variable indicates two things: 1) If negative, the variable indicates that the PMO is detached. 2) The variable is used to determine whether a thread invoking detach has permission to do so.

\paragraph{detach}
When an application invokes detach, the kernel searches for the PMO associated with the running process and name by searching for the VMA associated with the specified address and process. If the kernel fails to find one, then the detach attempt fails; otherwise, if the kernel succeeds in finding an attached PMO, then detach will set \verb attached_pid  to $0$. It will then traverse the linked list described in Section \ref{pagefaulthandler}, changing the permissions on each page in the linked list to \verb PROT_NONE , and finally free the node. 

\paragraph{psync}
As we described in Section \ref{sec:design},  psync forces any modifications to the PMO to be durable, by copying the data from the shadow PMO into the primary PMO. The function, which copies the data, also flushes the cache lines at the same time, with \texttt{memcpy\_flushcache}.

To optimize the synchronization of larger PMOs while maintaining atomicity, psync {\em walks} the page table for pages associated with the attached PMO. It checks each page table entry (PTE) for the soft-dirty bit, in a CRIU-like (Checkpoint/Restore In Userspace) manner\footnote{The central difference is that CRIU checkpoints these changes to block storage, whereas psync writes these changes from the shadow PMO to the primary PMO and flushes the associated cache lines, making psync crash consistent.}~\cite{emelyanovsoftdirty}. Modified pages' indices are added to a linked list, and all associated cache lines are flushed to persist the page. After all pages have been persisted (and their dirty bits cleared), a memory barrier is emitted. The kernel then traverses the linked list, copying each page from the shadow copy to the primary copy using memcpy\_flushcache. 

As we described in the design, the only possible way to handle multiple threads operating on a PMO undergoing a psync is to treat it as undefined behavior. Therefore, we assume the programmer will use a barrier such as OpenMP to ensure that no thread may write to, or invoke psync on, the PMO during the synchronization process.

\paragraph{PMO Recovery and crash consistency}
When an attach is called, the kernel checks the state of the PMO to be attached, as described in Section \ref{sec:recovery}. Recovery takes a linear $O(n)$ time; the total time for recovery is the time it takes to copy from the shadow PMO to the primary PMO, or vice-versa, so it is bound to the size of the PMO. In the case where the PMO was in the Detached state, or was attached in the Reading state, recovery takes constant $O(1)$ time.

We verify that our implementation of PMOs ensures crash consistency by modifying the psync system call and inserting a \texttt{panic} call immediately after the persist stage, but before the copy stage, in order to generate a kernel panic, and force the system to crash. When the system is restarted, we reattach the PMO, and verify that the data from the previous psync are in the PMO, as expected. We also test inserting panic immediately before the persist stage; after restart, the data from before the psync are there, which is expected behavior. Therefore, our implementation ensures both PMO crash consistency and recovery.

\section{Evaluation}
\label{sec:evaluation}

\paragraph{Evaluated Schemes}
In order to test our PMO system performance, we compare it against to two schemes. The baseline scheme represents the ``performance-ideal'' scheme where no crash consistency is provided. For this ideal Non-Crash-Consistent (NCC) scheme, we use the ext4-dax filesystem as the ideal baseline that uses Intel libpmem's pmem\_persist to persist updates but without any crash consistency. A crash with NCC will likely cause data corruption with dangling or invalid pointers, from which the original data structure may be unrecoverable. We would like to see how much overhead our scheme introduces against this performance-ideal NCC. The second scheme we compare our scheme against is the state of the art crash-consistent filesystem-based alternative, NOVA-Fortis~\cite{jian2017novafortis}, that employs snapshot to support crash-consistency. We note that NOVA-Fortis only guarantees crash consistency of the file system but does not guarantee crash consistency of application data, from the application point of view. We compare our PMO system against NCC and NOVA-Fortis in terms of execution time of a workload and I/O bandwidth. 

\paragraph{System Testbed}
The system we used for our evaluation is described in Table \ref{tab:system}. Each socket has two 32GiB DRAM DIMMs plus two 128GiB PMEM DIMMs. We use a stock version of the 5.9.13 Linux kernel for evaluating NCC  ext4 with dax. We implemented our PMO by modifying that version of the kernel. We use version 5.1.0 of the Linux kernel for NOVA-Fortis, as it is the latest version of the kernel NOVA-Fortis supports.

\begin{table}[htbp]
\centering
\footnotesize
\caption{System used for evaluation.}
    \label{EvalSys}
\begin{tabular}{|c|p{5.5cm}|}

\hline
\rowcolor[HTML]{EFEFEF} 
\textbf{Component} & \textbf{Specifications}  \\ \hline
Motherboard        & Dual socket Supermicro X11DPi-NT  (w/ADR)  \\ \hline
CPU                & $2\times$Intel Xeon Gold 6230, 20 cores, 40 threads \\ \hline
CPU Clock          & $2.1$GHz ($3.9$GHz Boost)   \\ \hline
Cache              & L1: 32KiB; L2: 1MiB; L3: 27.5MiB  \\ \hline
DRAM               & $4\times32$GiB DDR4 @ $2666$MHz  \\ \hline
PMEM               & $4\times128$GiB Intel Optane DC \\ \hline
Kernel             & Linux 5.9.13 (PMO, NCC), 5.1.0 (NOVA) \\ \hline
OS   & $64$ Bit Fedora $33$      \\ \hline
\end{tabular}
\label{tab:system}
\end{table}
\definecolor{darkgreen}{rgb}{0.0, 0.2, 0.13}
\definecolor{peach}{HTML}{FFE1A8}
\definecolor{terracotta}{HTML}{E26D5C}
\definecolor{darkelectricblue}{HTML}{53687E}
\definecolor{grannysmithapple}{HTML}{BDF7B7}
\definecolor{orangeyellow}{HTML}{F5B700}
\definecolor{darkcyan}{HTML}{297373}

\usetikzlibrary{patterns}

\pgfplotstableread[col sep=comma, row sep=newline]
    {figures/results/perfavg/PMO.csv}\PMO
\pgfplotstableread[col sep=comma, row sep=newline]
    {figures/results/perfavg/PMEM.csv}\PMEM
\pgfplotstableread[col sep=comma, row sep=newline]
    {figures/results/perfavg/NOVA.csv}\NOVA

\pgfplotstableread[col sep=comma, row sep=newline]
    {figures/results/bandwidth/bandPMO.csv}\bandPMO
\pgfplotstableread[col sep=comma, row sep=newline]
    {figures/results/bandwidth/bandPMEM.csv}\bandPMEM
\pgfplotstableread[col sep=comma, row sep=newline]
    {figures/results/bandwidth/bandNOVA.csv}\bandNOVA

\pgfplotstableread[col sep=comma, row sep=newline]
    {figures/results/threads/PMO/LU.csv}\LUPMOT
\pgfplotstableread[col sep=comma, row sep=newline]
    {figures/results/threads/PMO/2D.csv}\TWODPMOT
\pgfplotstableread[col sep=comma, row sep=newline]
    {figures/results/threads/PMO/TMM.csv}\TMMPMOT

\pgfplotstableread[col sep=comma, row sep=newline]
    {figures/results/threads/NOVA/LU.csv}\LUNOVAT
\pgfplotstableread[col sep=comma, row sep=newline]
    {figures/results/threads/NOVA/2D.csv}\TWODNOVAT
\pgfplotstableread[col sep=comma, row sep=newline]
    {figures/results/threads/NOVA/TMM.csv}\TMMNOVAT

\pgfplotstableread[col sep=comma, row sep=newline]
    {figures/results/sensitivity/PMO/LU.csv}\LUPMOS
\pgfplotstableread[col sep=comma, row sep=newline]
    {figures/results/sensitivity/PMO/2D.csv}\TWODPMOS
\pgfplotstableread[col sep=comma, row sep=newline]
    {figures/results/sensitivity/PMO/TMM.csv}\TMMPMOS

\pgfplotstableread[col sep=comma, row sep=newline]
    {figures/results/sensitivity/NOVA/LU.csv}\LUNOVAS
\pgfplotstableread[col sep=comma, row sep=newline]
    {figures/results/sensitivity/NOVA/2D.csv}\TWODNOVAS
\pgfplotstableread[col sep=comma, row sep=newline]
    {figures/results/sensitivity/NOVA/TMM.csv}\TMMNOVAS

\definecolor{vermillion}{HTML}{D64550}
\definecolor{emerald}{HTML}{32936F}
\definecolor{lapis}{HTML}{255C99}
\definecolor{flax}{HTML}{E3D081}
\definecolor{lavender}{HTML}{CBBAED}

\definecolor{darkgreen}{rgb}{0.0, 0.2, 0.13}
\definecolor{peach}{HTML}{FFE1A8}
\definecolor{terracotta}{HTML}{E26D5C}
\definecolor{darkelectricblue}{HTML}{53687E}
\definecolor{grannysmithapple}{HTML}{BDF7B7}
\definecolor{orangeyellow}{HTML}{F5B700}
\definecolor{darkcyan}{HTML}{297373}

\paragraph{Benchmarks}

In order to measure PMO performance from the perspective of {\em execution time}, we use an OpenMP version of LU decomposition provided by~\cite{LU}, (and originally from the SPLASH benchmark suite~\cite{Sakalis2016splash3}), a 2D-Convolution (2dConv) benchmark and a Tiled Matrix Multiplication (TMM) benchmark, taken from a recent persistent memory study~\cite{Elnawawy-PACT17-checkpointing}. We run LU, 2dConv and TMM with matrix sizes $3584\times3584$ doubles, $4096\times128$ integers, and $3072\times3072$ integers, respectively.

We ported these benchmarks by replacing their dynamic memory allocation calls (e.g., malloc and calloc) with a pair of pcreate and attach (to create and map into the process'  address space a PMO of required size) and with  pmem\_map\_file for NOVA-Fortis. Each benchmark ported to PMO system uses multiple PMOs determined by number of memory allocation calls in original version. At the end of each iteration of the performance critical loop in the benchmarks, if a specified time duration $\Delta$ has elapsed from the previous invocation of the synchronization, we insert a synchronization point (i.e., psync in case of PMO, generating a snapshot in case of NOVA-Fortis, and invoking pmem\_persist in case of ext4-dax (i.e., NCC)). In this way, by varying $\Delta$ we can vary the synchronization rate. Note that pmem\_persist persists updates but without crash consistency.

\begin{table}[htbp]
\footnotesize
\centering
    \caption{Filebench workloads.}
    \label{tab:Filebench}
\begin{tabular}{|c|c|c|}
\hline
\rowcolor[HTML]{EFEFEF} 
\textbf{Workload} & \textbf{Description} & \textbf{Wr\%} \\ \hline
FileServer (FS) & File Server  & 67\% \\ \hline
VarMail (VM)  & Mail Server  & 50\% \\ \hline
WebProxy (WP)  & Proxy Server  & 16\% \\ \hline
WebServer (WS) & Web Server &  9\% \\ \hline
\end{tabular}
\end{table}

For measuring I/O bandwidth performance, we rely on  FileBench benchmarks~\cite{tarasov2016filebench} representing I/O intensive real-world applications.  We ported these benchmarks to PMO-system by replacing files with PMOs of respective sizes. For ext4-dax (i.e., NCC) and NOVA-Fortis we mapped files via DAX. Furthermore, we insert synchronization points in the benchmarks after \textbf{every} update (i.e., \textit{append} or \textit{wholefilewrite} flowop in terminology of FileBench). Also, to avoid races between threads, we emit a pthread barrier before and after each psync. Table \ref{tab:Filebench} shows the FileBench benchmarks and their characteristics.

\subsection{Performance Evaluation}

For our evaluation, we want to answer several questions: How much performance overhead does our PMO system incur in comparison with a system with no crash consistency (NCC), and the state-of-the-art file system approach NOVA-fortis? How scalable is our PMO system, as the number of threads scale, and as the frequency of psync increases? 

\begin{figure}[htbp]
\centering
\begin{tikzpicture}

    \begin{axis}[ybar,height=1.6in,width=3in,
            ymin=0,ymax=1.1,bar width=7pt,
            symbolic x coords={LU,2D,TMM,GMEAN},
            legend style={at={(0.5,1.2)},anchor=north, legend columns =-1},
            ylabel style={align=center},
            ylabel=Speedup ratio
        ]
        
        \addplot    [black,
                    pattern=north east lines,
                    pattern color=black,
                   preaction={fill=flax!100!white}]
                   table[x=benchmark,y=runtime]{\PMEM};
                    
        \addplot    [black,
                    pattern=north west lines,
                    pattern color=grannysmithapple,
                    preaction={fill=darkelectricblue!100!white}]
                    table[x=benchmark,y=runtime]{\PMO};
                    
        \addplot    [black,
                    pattern=crosshatch dots,
                    pattern color=black,
                    preaction={fill=vermillion!100!white}]
                    table[x=benchmark,y=runtime]{\NOVA};

        \legend{NCC, PMO, NOVA-Fortis}
    \end{axis}
\end{tikzpicture}
 \caption{Performance comparison of crash consistent PMO and NOVA-Fortis to Non-Crash-Consistent (NCC) ext4-dax.}
 \label{fig:par}
\end{figure}
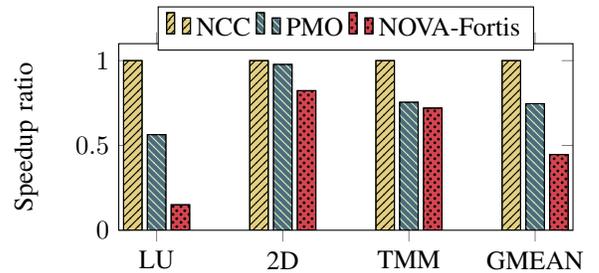

Figure \ref{fig:par} compares the performance of crash consistent systems i.e., PMO and NOVA-Fortis with the Non-Crash-Consistent (NCC) system i.e., ext4-dax. Results are normalized to NCC and reported for benchmarks executed with 16 threads while synchronization is performed at rate of four times per second. When compared with NCC, our PMO system slows down the execution time by only 25.4\% (geometric mean) vs. 55.36\% with NOVA-Fortis. Thus, our PMO system is not far off from NCC, and is $\approx 1.67\times$ faster than NOVA-Fortis.

This can be attributed to the fact that, unlike NCC, crash-consistent systems employ additional mechanisms to support crash-consistency i.e., shadowing in PMO-system and snapshot in NOVA-Fortis. Our PMO system performs better as it synchronizes, at each synchronization point in a benchmark, only those PMOs that are actively used by the benchmark. On the other hand, NOVA-Fortis takes a crash-consistent image of the whole filesystem at each synchronization point and not only the files that are in active use. This also illustrates the strength of an application-centric approach for crash-consistency.

\begin{figure}[htbp]
\centering
\begin{tikzpicture}
    \begin{axis}[ybar,height=1.6in,width=3in,
            ybar,            
            xtick distance=1,
            ymin=0.0,
            ymax=1.1,bar width=7pt,
            symbolic x coords={FS, WS, WP, VM, GM},
            legend style={at={(0.5,1.2)},anchor=north, legend columns =-1},
            ylabel style={align=center},
            ylabel=Bandwidth \\ (normalized)
            ]
                    \usetikzlibrary{patterns}

        \addplot    [black,
                    pattern=north east lines,
                    pattern color=black,
                   preaction={fill=flax!100!white}] table[x=benchmark,y=runtime]{\bandPMEM};
                    
        \addplot    [black,
                    pattern=north west lines,
                    pattern color=grannysmithapple,
                    preaction={fill=darkelectricblue!100!white}]
                    table[x=benchmark,y=runtime]{\bandPMO};
                    
        \addplot    [black,
                    pattern=crosshatch dots,
                    pattern color=black,
                    preaction={fill=vermillion!100!white}]
                    table[x=benchmark,y=runtime]{\bandNOVA};
        \legend{NCC, PMO, NOVA-Fortis}
    \end{axis}
    \end{tikzpicture}
\label{fig:filebench}
 \caption{Bandwidth comparison of crash consistent PMO and NOVA-Fortis to Non-Crash-Consistent (NCC) ext4-dax.}
\label{fig:filebench}
\end{figure}
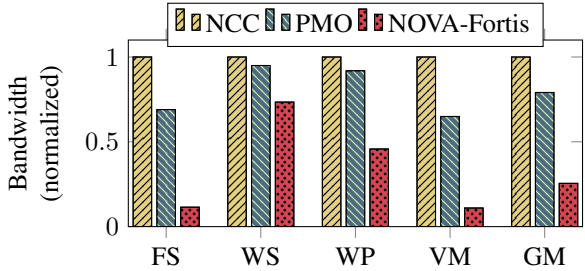

Figure \ref{fig:filebench} compares the I/O bandwidth achieved by our PMO system compared to NCC and NOVA-fortis. Results are normalized to NCC and reported for 16 threads with synchronization performed on every update operation. On average, shown by geometric mean (GM) bar, PMO system and Nova-fortis provide crash-consistency at the expense of losing $21\%$ and $74.5\%$ bandwidth, respectively. This result means that our PMO system achieves bandwidth $3\times$ higher than NOVA-Fortis. Performance of both PMO-system and NOVA-Fortis, vary across benchmarks as each benchmark has a different number of synchronization points in accordance with their write percentage (Table \ref{tab:Filebench}). More frequent synchronization incurs more overhead and hence the lower bandwidth performance.

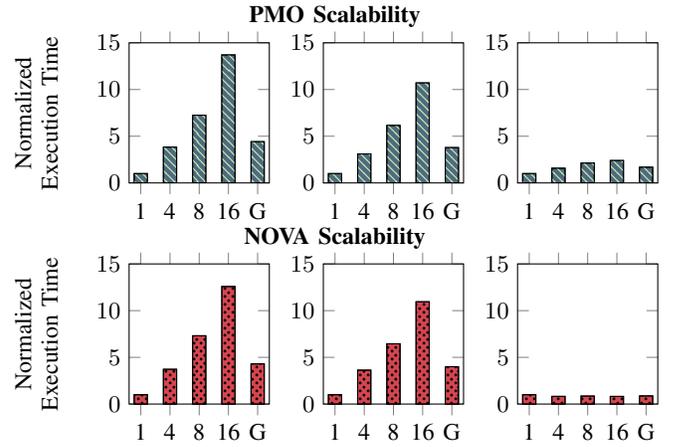
\begin{figure}[h]
\hfill
  \centering
\begin{subfigure}[b]{0.5\textwidth}
\centering \small\textbf{PMO Scalability}\vfill
          \begin{tikzpicture}
\begin{axis}[ybar,height=0.38\linewidth,width=0.38\linewidth,xtick distance=1,enlarge x limits=0.1,
 legend style={at={(0.5,1.3)},anchor=north,legend columns =-1},
  symbolic x coords={1,4,8,16,32},
    xticklabels = {1, 1,4,8,16,G},
            ymin=0,ymax=15,
            bar width=5pt,
            ylabel style={align=center},
            ylabel=Normalized \\ Execution Time ]
        \addplot    [black,
                    pattern=north west lines,
                    pattern color=grannysmithapple,
                    preaction={fill=darkelectricblue!100!white}]                   table[x=threads,y=exec]{\TWODPMOT};

\end{axis}
\end{tikzpicture}
          \begin{tikzpicture}
\begin{axis}[ybar,height=0.38\linewidth,width=0.38\linewidth,xtick distance=1,enlarge x limits=0.1,
 legend style={at={(0.5,1.3)},anchor=north,legend columns =-1},
  symbolic x coords={1,4,8,16,32},
    xticklabels = {1, 1,4,8,16,G},
            ymin=0,ymax=15,
            bar width=5pt]
        \addplot    [black,
                    pattern=north west lines,
                    pattern color=grannysmithapple,
                    preaction={fill=darkelectricblue!100!white}]
                   table[x=threads,y=exec]{\TMMPMOT};

\end{axis}
\end{tikzpicture}
\begin{tikzpicture}
\begin{axis}[ybar,height=0.38\linewidth,width=0.38\linewidth,xtick distance=1,enlarge x limits=0.1,
 legend style={at={(0.5,1.3)},anchor=north,legend columns =-1},
  symbolic x coords={1,4,8,16,32},
    xticklabels = {1, 1,4,8,16,G},
            ymin=0,ymax=15,
            bar width=5pt]
        \addplot    [black,
                    pattern=north west lines,
                    pattern color=grannysmithapple,
                    preaction={fill=darkelectricblue!100!white}]
                   table[x=threads,y=exec]{\LUPMOT};

\end{axis}
\end{tikzpicture}
\end{subfigure}
\hfill
\begin{subfigure}[b]{0.5\textwidth}
\centering \small\textbf{NOVA Scalability}
\vfill
  \centering
          \begin{tikzpicture}
\begin{axis}[ybar,height=0.38\linewidth,width=0.38\linewidth,xtick distance=1,enlarge x limits=0.1,
 legend style={at={(0.5,1.3)},anchor=north,legend columns =-1},
  symbolic x coords={1,4,8,16,32},
    xticklabels = {1, 1,4,8,16,G},
            ymin=0,ymax=15,
            bar width=5pt,
            ylabel style={align=center},
            ylabel=Normalized \\ Execution Time ]
        \addplot    [black,
                    pattern=crosshatch dots,
                    pattern color=black,
                    preaction={fill=vermillion!100!white}]
                   table[x=threads,y=exec]{\TWODNOVAT};

\end{axis}
\end{tikzpicture}
          \begin{tikzpicture}
\begin{axis}[ybar,height=0.38\linewidth,width=0.38\linewidth,xtick distance=1,enlarge x limits=0.1,
 legend style={at={(0.5,1.3)},anchor=north,legend columns =-1},
  symbolic x coords={1,4,8,16,32},
    xticklabels = {1, 1,4,8,16,G},
            ymin=0,ymax=15,
            bar width=5pt]
        \addplot    [black,
                    pattern=crosshatch dots,
                    pattern color=black,
                    preaction={fill=vermillion!100!white}]
                   table[x=threads,y=exec]{\TMMNOVAT};

\end{axis}
\end{tikzpicture}
\begin{tikzpicture}
\begin{axis}[ybar,height=0.38\linewidth,width=0.38\linewidth,xtick distance=1,enlarge x limits=0.1,
 legend style={at={(0.5,1.3)},anchor=north,legend columns =-1},
  symbolic x coords={1,4,8,16,32},
    xticklabels = {1, 1,4,8,16,G},
            ymin=0,ymax=15,
            bar width=5pt]
        \addplot    [black,
                    pattern=crosshatch dots,
                    pattern color=black,
                    preaction={fill=vermillion!100!white}]
                   table[x=threads,y=exec]{\LUNOVAT};

\end{axis}
\end{tikzpicture}
\end{subfigure}
\caption{Thread-scalability of PMO-system and NOVA-Fortis. From left to right: 2DConv, TMM, and LU.} 
\label{fig:scalability}
\end{figure}

Figure \ref{fig:scalability} shows execution-time of LU, 2DConv and TMM workloads when executed with  ${N(=1, 4, 8, 16)}$ thread and synchronized four times per second. Results are normalized to 1-thread execution while G-bar shows geometric mean. Results show that performance of PMO system scales better than that of NOVA-Fortis. However, the rate of performance scaling varies across three workloads in following increasing order: 2DConv, TMM, and LU. This is explained by number of pages to be updated per synchronization operation (i.e work assigned to threads) in each workload. For example, 2Dconv, TMM and LU copy 184, 9216, and 24451 pages per psync from shadow to primary PMO,  respectively. 

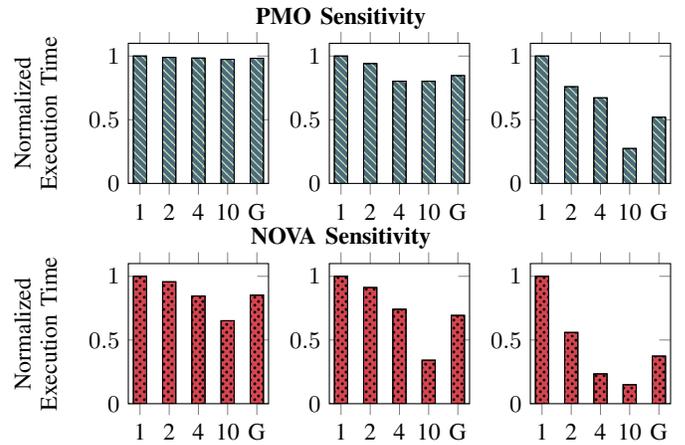
\begin{figure}[h]
\hfill
  \centering
\begin{subfigure}[b]{0.5\textwidth}
\centering \small \textbf{PMO Sensitivity}\vfill
          \begin{tikzpicture}
\begin{axis}[ybar,height=0.38\linewidth,width=0.38\linewidth,xtick distance=1,enlarge x limits=0.1,
 legend style={at={(0.5,1.3)},anchor=north,legend columns =-1},
  symbolic x coords={1,2,4,10,11},
    xticklabels = {1, 1,2,4,10,G},
            ymin=0,ymax=1.1,
            bar width=5pt,
            ylabel style={align=center},
            ylabel=Normalized \\ Execution Time 
            ]
        \addplot    [black,
                    pattern=north west lines,
                    pattern color=grannysmithapple,
                    preaction={fill=darkelectricblue!100!white}]
                   table[x=threads,y=exec]{\TWODPMOS};

\end{axis}
\end{tikzpicture}
          \begin{tikzpicture}
\begin{axis}[ybar,height=0.38\linewidth,width=0.38\linewidth,xtick distance=1,enlarge x limits=0.1,
 legend style={at={(0.5,1.3)},anchor=north,legend columns =-1},
  symbolic x coords={1,2,4,10,11},
    xticklabels = {1, 1,2,4,10,G},
            ymin=0,ymax=1.1,
            bar width=5pt]
        \addplot    [black,
                    pattern=north west lines,
                    pattern color=grannysmithapple,
                    preaction={fill=darkelectricblue!100!white}]
                   table[x=threads,y=exec]{\TMMPMOS};

\end{axis}
\end{tikzpicture}
\begin{tikzpicture}
\begin{axis}[ybar,height=0.38\linewidth,width=0.38\linewidth,xtick distance=1,enlarge x limits=0.1,
 legend style={at={(0.5,1.3)},anchor=north,legend columns =-1},
  symbolic x coords={1,2,4,10,11},
    xticklabels = {1, 1,2,4,10,G},
            ymin=0,ymax=1.1,
            bar width=5pt]
        \addplot    [black,
                    pattern=north west lines,
                    pattern color=grannysmithapple,
                    preaction={fill=darkelectricblue!100!white}]
                   table[x=threads,y=exec]{\LUPMOS};

\end{axis}
\end{tikzpicture}
\end{subfigure}
\hfill
\begin{subfigure}[b]{0.5\textwidth}
\centering \small \textbf{NOVA Sensitivity}
\vfill
  \centering
          \begin{tikzpicture}
\begin{axis}[ybar,height=0.38\linewidth,width=0.38\linewidth,xtick distance=1,enlarge x limits=0.1,
 legend style={at={(0.5,1.3)},anchor=north,legend columns =-1},
  symbolic x coords={1,2,4,10,11},
    xticklabels = {1, 1,2,4,10,G},
            ymin=0,ymax=1.1,
            bar width=5pt,
            ylabel style={align=center},
            ylabel=Normalized \\ Execution Time 
            ]
        \addplot    [black,
                    pattern=crosshatch dots,
                    pattern color=black,
                    preaction={fill=vermillion!100!white}]
                   table[x=threads,y=exec]{\TWODNOVAS};

\end{axis}
\end{tikzpicture}
          \begin{tikzpicture}
\begin{axis}[ybar,height=0.38\linewidth,width=0.38\linewidth,xtick distance=1,enlarge x limits=0.1,
 legend style={at={(0.5,1.3)},anchor=north,legend columns =-1},
  symbolic x coords={1,2,4,10,11},
    xticklabels = {1, 1,2,4,10,G},
            ymin=0,ymax=1.1,
            bar width=5pt]
        \addplot    [black,
                    pattern=crosshatch dots,
                    pattern color=black,
                    preaction={fill=vermillion!100!white}]
                   table[x=threads,y=exec]{\TMMNOVAS};

\end{axis}
\end{tikzpicture}
\begin{tikzpicture}
\begin{axis}[ybar,height=0.38\linewidth,width=0.38\linewidth,xtick distance=1,enlarge x limits=0.1,
 legend style={at={(0.5,1.3)},anchor=north,legend columns =-1},
  symbolic x coords={1,2,4,10,11},
    xticklabels = {1, 1,2,4,10,G},
            ymin=0,ymax=1.1,
            bar width=5pt]
        \addplot    [black,
                    pattern=crosshatch dots,
                    pattern color=black,
                    preaction={fill=vermillion!100!white}]
                   table[x=threads,y=exec]{\LUNOVAS};

\end{axis}
\end{tikzpicture}
\end{subfigure}
\caption{Synchronization-sensitivity of PMO-system and NOVA-Fortis. From left to right: 2DConv, TMM, and LU. } 
\label{fig:sensitivity}
\end{figure}

Figure \ref{fig:sensitivity} shows normalized execution-time of LU, 2DConv and TMM workloads executed with 16 threads and ${N(=1, 2, 4, 10)}$ synchronization points inserted per second (i.e., psync for PMO and snapshot for NOVA-Fortis).
Dues to its application-centric  crash-consistency approach, PMO system shows less sensitivity to synchronization-rate than NOVA-Fortis for all three workloads.

\section{Conclusion}\label{sec:conclusions}

This paper has described a novel design of the PMO abstraction, its system properties, intuitive programming interface, and semantics.
PMOs are designed to be more efficient than DAX enabled files, crash-consistent, and provide easy to use semantics for persistency and crash-consistency. 
We discussed an implementation of a persistent memory object system for the Linux kernel in real hardware utilizing Intel Optane DC Memory. We demonstrated that for crash-consistent updates, PMO system has relatively lower overhead over a non-crash-consistent ext4-dax system and performs $1.67\times$ and $3\times$ faster than  NOVA-Fortis, for two sets of evaluated benchmarks. It is also found to be more thread-scalable and less sensitive to synchronization rate than NOVA-Fortis.

\bibliographystyle{plain}
\bibliography{main}

\vspace{12pt}

\end{document}